\documentclass[twocolumn,aps,prc,longbibliography,superscriptaddress,nofootinbib,showkeys, floatfix]{revtex4-2}

\usepackage[colorlinks=true,linkcolor=blue,citecolor=blue]{hyperref}
\usepackage[utf8]{inputenc}
\usepackage{hyperref}
\usepackage{xcolor}
\usepackage{slashed,indentfirst}
\usepackage{amsmath}
\usepackage[symbol,splitrule]{footmisc}
\usepackage{babel,csquotes,xpatch}
\usepackage{dcolumn,multirow}
\usepackage{notoccite}
\usepackage{graphicx,color,epstopdf}
\usepackage{subfigure}
\usepackage{subfiles}
\usepackage{bm}
\usepackage{tabularx}
\usepackage{natbib}
\usepackage{float}
\usepackage{xeCJK}
\usepackage{mathptmx}
\usepackage{txfonts}
\usepackage{comment}
\usepackage{autobreak}
\usepackage{layouts}

\hypersetup{
    colorlinks=true,
    linkcolor=blue,
    citecolor=blue,
    urlcolor=blue,
    filecolor=magenta
}

\begin{document}
\preprint{AIP/123-QED}

\newcommand{\nuc}[2]{$^{#1}$#2}
\newcommand{\degree}{\,$^\circ$ }

\title{Isoscalar Giant Resonances in the even-$A$ Pd Isotopes}

\author{J. Arroyo}
  \email{jarroyo1@nd.edu}
\affiliation{ Department of Physics and Astronomy, University of Notre Dame, Notre Dame, IN 46556, USA
}
\author{U. Garg}
 \email{garg@nd.edu}
\affiliation{ Department of Physics and Astronomy, University of Notre Dame, Notre Dame, IN 46556, USA
}
\author{T. Furuno}
\affiliation{Department of Physics, The University of Osaka, 1-1 Machikaneyama, Toyonaka, Osaka 560-0043, Japan
}
\affiliation{
Department of Applied Physics, University of Fukui, 3-9-1 Bunkyo, Fukui, Fukui 910-8507, Japan
}
\author{M. Itoh}
\affiliation{Research Center for Accelerator and Radioisotope Science (RARiS), Tohoku University, 6-3 Aoba, Aramaki, Aoba, Sendai, Miyagi 980-8578, Japan
}
\author{H. Shimojo}
\affiliation{Department of Physics, The University of Osaka, 1-1 Machikaneyama, Toyonaka, Osaka 560-0043, Japan
}
\author{S. Adachi}
\affiliation{Research Center for Accelerator and Radioisotope Science (RARiS), Tohoku University, 6-3 Aoba, Aramaki, Aoba, Sendai, Miyagi 980-8578, Japan
}
\author{H. Akimune}
\affiliation{Department of Physics, Konan University, 8-9-1 Okamoto, Higashinada, Kobe, Hyogo 658-8501, Japan
}
\author{J. Cai}
\affiliation{Research Center for Nuclear Physics (RCNP), The University of Osaka, 10-1 Mihogaoka, Ibaraki, Osaka 567-0047, Japan
}
\author{G.~Colò}
\affiliation{Dipartimento di Fisica, Universit\`a degli Studi di Milano, via Celoria 16, 20133 Milano, Italy}
\affiliation{INFN sezione di Milano, via Celoria 16, 20133 Milano, Italy
}
\author{M.~Dozono}
\affiliation{Department of Physics, Kyoto University, Kitashirakawa-Oiwake, Sakyo, Kyoto 606-8502, Japan
}
\author{F. Endo}
\affiliation{Research Center for Nuclear Physics (RCNP), The University of Osaka, 10-1 Mihogaoka, Ibaraki, Osaka 567-0047, Japan
}
\author{M. Fujiwara}
\affiliation{Research Center for Nuclear Physics (RCNP), The University of Osaka, 10-1 Mihogaoka, Ibaraki, Osaka 567-0047, Japan
}
\author{F. Furukawa}
\affiliation{Research Center for Nuclear Physics (RCNP), The University of Osaka, 10-1 Mihogaoka, Ibaraki, Osaka 567-0047, Japan
}
\author{M.N. Harakeh}
\affiliation{Nuclear Energy Group, ESRIG, University of Groningen, 9747 AA Groningen, The Netherlands
}
\author{Y.~Hijikata}
\affiliation{Department of Physics, Kyoto University, Kitashirakawa-Oiwake, Sakyo, Kyoto 606-8502, Japan
}
\author{Y.~Honda}
\affiliation{Department of Physics, The University of Osaka, 1-1 Machikaneyama, Toyonaka, Osaka 560-0043, Japan
}
\author{G.~Hosoya}
\affiliation{Research Center for Accelerator and Radioisotope Science (RARiS), Tohoku University, 6-3 Aoba, Aramaki, Aoba, Sendai, Miyagi 980-8578, Japan
}
\author{N. Itakura}
\affiliation{Research Center for Nuclear Physics (RCNP), The University of Osaka, 10-1 Mihogaoka, Ibaraki, Osaka 567-0047, Japan
}
\author{K. Kawata}
\affiliation{Research Center for Nuclear Physics (RCNP), The University of Osaka, 10-1 Mihogaoka, Ibaraki, Osaka 567-0047, Japan
}
\author{T. Kawabata}
\affiliation{Department of Physics, The University of Osaka, 1-1 Machikaneyama, Toyonaka, Osaka 560-0043, Japan
}
\author{N. Kobayashi}
\affiliation{Research Center for Nuclear Physics (RCNP), The University of Osaka, 10-1 Mihogaoka, Ibaraki, Osaka 567-0047, Japan
} 
\author{Z.Z.~Li~(李~征~征)}
\affiliation{Center for Computational Sciences, University of Tsukuba, Tsukuba 305-8577, Japan}

\affiliation{State Key Laboratory of Nuclear Physics and Technology, School of Physics, Peking University, Beijing 100871, China}

\author{Y.~Lin}
\affiliation{Department of Physics, The University of Osaka, 1-1 Machikaneyama, Toyonaka, Osaka 560-0043, Japan
}
\author{Y. Matsuda}
\affiliation{Research Center for Nuclear Physics (RCNP), The University of Osaka, 10-1 Mihogaoka, Ibaraki, Osaka 567-0047, Japan
}
\author{T. Morishita}
\affiliation{Department of Physics, Konan University, 8-9-1 Okamoto, Higashinada, Kobe, Hyogo 658-8501, Japan
}
\author{K. Nakano}
\affiliation{Department of Physics, Konan University, 8-9-1 Okamoto, Higashinada, Kobe, Hyogo 658-8501, Japan
}
\author{Y.F.~Niu~(牛一斐)}
\affiliation{School of Physics and Astronomy, Shanghai Jiao Tong University,
Key Laboratory for Particle Astrophysics and Cosmology (MoE), Shanghai 200240, China}

\author{T.~Okamura}
\affiliation{Department of Physics, The University of Osaka, 1-1 Machikaneyama, Toyonaka, Osaka 560-0043, Japan
}
\author{S.~Ota}
\affiliation{Research Center for Nuclear Physics (RCNP), The University of Osaka, 10-1 Mihogaoka, Ibaraki, Osaka 567-0047, Japan
}
\author{F. Saito}
\affiliation{Department of Physics, Konan University, 8-9-1 Okamoto, Higashinada, Kobe, Hyogo 658-8501, Japan
}
\author{R. Saito}
\affiliation{Research Center for Accelerator and Radioisotope Science (RARiS), Tohoku University, 6-3 Aoba, Aramaki, Aoba, Sendai, Miyagi 980-8578, Japan
}
\author{S. Sakajo}
\affiliation{Department of Physics, The University of Osaka, 1-1 Machikaneyama, Toyonaka, Osaka 560-0043, Japan
}
\author{K. Sakanashi}
\affiliation{Department of Physics, The University of Osaka, 1-1 Machikaneyama, Toyonaka, Osaka 560-0043, Japan
}
\author{H. Shibakita}
\affiliation{Research Center for Nuclear Physics (RCNP), The University of Osaka, 10-1 Mihogaoka, Ibaraki, Osaka 567-0047, Japan
}
\author{R.~Tsuji}
\affiliation{Department of Physics, Kyoto University, Kitashirakawa-Oiwake, Sakyo, Kyoto 606-8502, Japan
}
\author{G.~Umemoto}
\affiliation{Department of Physics, Konan University, 8-9-1 Okamoto, Higashinada, Kobe, Hyogo 658-8501, Japan
}
\author{A. Yamasaki}
\affiliation{Department of Physics, Konan University, 8-9-1 Okamoto, Higashinada, Kobe, Hyogo 658-8501, Japan
}
\author{S. Yamazaki}
\affiliation{Research Center for Accelerator and Radioisotope Science (RARiS), Tohoku University, 6-3 Aoba, Aramaki, Aoba, Sendai, Miyagi 980-8578, Japan
}
\author{T. Yano}
\affiliation{Department of Physics, Kyoto University, Kitashirakawa-Oiwake, Sakyo, Kyoto 606-8502, Japan
}
\author{K. Yasumura}
\affiliation{Department of Physics, Konan University, 8-9-1 Okamoto, Higashinada, Kobe, Hyogo 658-8501, Japan
}
\author{S. Yonekura}
\affiliation{Research Center for Accelerator and Radioisotope Science (RARiS), Tohoku University, 6-3 Aoba, Aramaki, Aoba, Sendai, Miyagi 980-8578, Japan
}
\author{J. Zenihiro}
\affiliation{Department of Physics, Kyoto University, Kitashirakawa-Oiwake, Sakyo, Kyoto 606-8502, Japan
}

\date{\today}

\begin{abstract}
    Studies of the isoscalar giant monopole resonance (ISGMR) across the chart of nuclides provide insight into the incompressibility of nuclear matter near saturation density, $K_\infty$. Such studies had revealed a discrepancy between theoretical approaches: quasiparticle random phase approximation (QRPA) derived from Skyrme interactions reproduce the strength distributions of the ISGMR in the ``doubly-closed-shell'' nuclei \nuc{90}{Zr} and \nuc{208}{Pb}, but their descriptions of strength distributions in open-shell medium-heavy nuclei suggest higher centroid energies should be experimentally observed. The latter nuclei required a smaller $K_{\infty}$ and were thus deemed ``softer''. The present work serves to add to this ``softness'' discourse by extracting ISGMR strength distributions for $^{104,106,108,110}$Pd via 386-\,MeV inelastic $\alpha$-scattering. The extracted giant resonance strength distributions are consistent with expectations in this isotopic range. Additional Quasiparticle Vibration Coupling (QPVC) effects are included with the QRPA approach and compared to aforementioned ISGMR strength distributions.
    
\end{abstract}


\maketitle

\section*{\label{sec:intro}I. Introduction}

Nuclear incompressibility ($K_{\infty}$) is a parameter of great importance in describing the stiffness of nuclear matter near saturation density. Experimental determination of $K_\infty$ enables a better understanding of systems modeled by a collection of bound nucleons, with far-reaching consequences in astrophysical systems or heavy-ion collisions \cite{BETHE1979487,Oertel2017,Bertsch1988,1992ApJ...398..569L}. Historically, the study of the isoscalar giant monopole resonance (ISGMR) has provided key insights into the nuclear equation of state, especially $K_\infty$, by cataloging finite compressibilities, $K_A$, for medium-heavy nuclei \cite{UGarg}. These values have been used as input to tune modern mean-field models to obtain predictions for $K_\infty$ \cite{PhysRevC.70.024307}.

\begin{figure*}[ht!]
    \centering
    \includegraphics[width=17.cm, height=6cm]{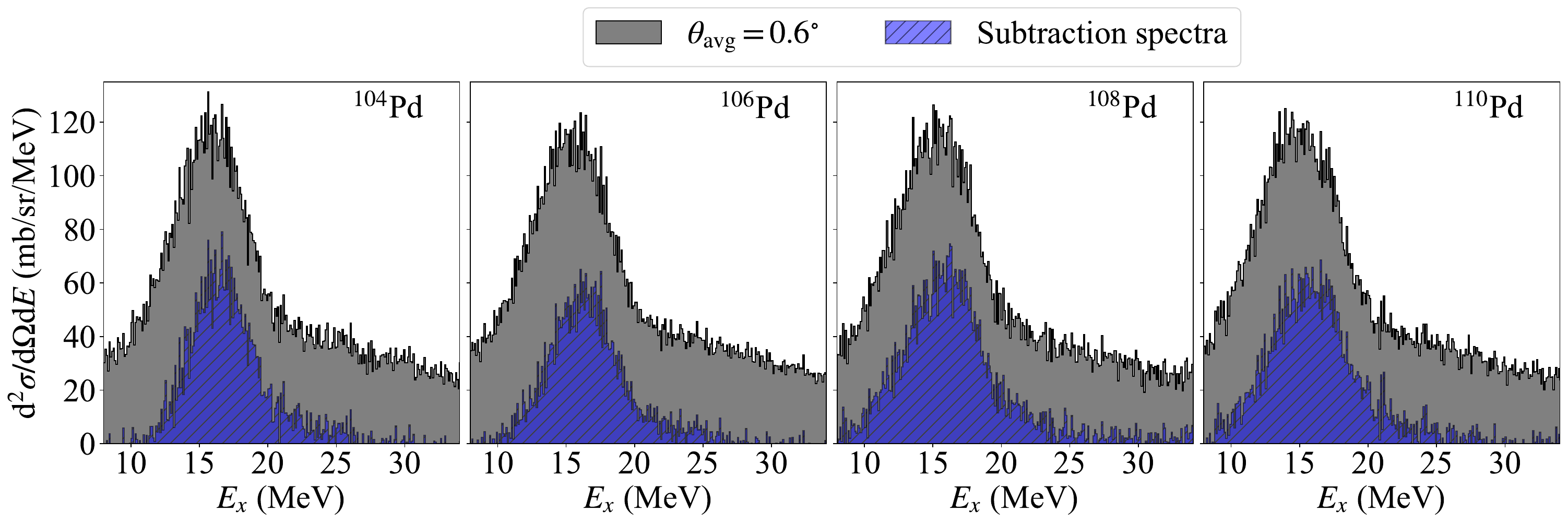}
    \caption{\label{fig:fwd_spectra}(Color online) Background-subtracted, inelastic scattering cross sections for all nuclei investigated in this work. Shown are the more forward spectra at 0.6\degree average spectrometer angle (grey shaded) and for each of these nuclei, the respective ``subtraction spectra'', exhibiting a strong ISGMR response in each case (see text).}
\end{figure*}

Experimental values for $K_\infty$ have been generally obtained from ISGMR data for well behaved ``doubly-closed-shell'' nuclei, $^{90}$Zr and $^{208}$Pb \cite{PhysRevC.70.024307, Shlomo2006}. However, with the advent of high-precision data for the tin isotopes~\cite{TaoLi_116Sn}, problems with this approach began to arise. Mean-field methods, such as Quasi-particle Random Phase Approximation (QRPA) based on Skyrme energy density functionals tuned on stiff nuclei, systematically overestimated ISGMR centroid energies for the tin isotopes, an effect first seen in Ref. \cite{TaoLi_116Sn}. This continued to be the case as more data became available, such as in the Cd and Mo isotopic chains \cite{PATEL2012447, KBHOWARD_moly}. This observed ``softness'' of open-shell nuclei had remained an outstanding problem in nuclear structure for over 20 years \cite{npa-garg}; plausible solutions to this problem have recently been advanced, however \cite{Z.Z.Li_2023_PRL_ISGMR,LITVINOVA2023}.

Extensive investigations carried out at facilities such as Kernfysisch Versneller Instituut (KVI) at the University of Groningen, the Cyclotron Institute at the Texas A\&M University, the Research Center for Nuclear Physics (RCNP) at the University of Osaka \cite{UGarg}, and more recently, the iThemba Lab in South Africa \cite{Bahini2024}, have led to a wealth of available data on the ISGMR across the chart of nuclides. This information continues to fuel theoretical endeavors to predict ISGMR parameters, leading to a range of values for $K_\infty$ depending on the particular parameter set used for those calculations. A response to this complex situation  has led to beyond mean-field attempts to better characterize the ISGMR and its link with the nuclear equation of state. As well, relativistic approaches have been applied to capture the complexity that may be the result of the aforementioned softness discrepancy \cite{npa-garg, piekarewicz1, piekarewicz2}. Therefore, it remains imperative to continue to provide high-quality ISGMR data to assist in resolving the contrast between the experimental results and the modeled nuclear matter equation of state.

This paper follows in the footsteps of its predecessors \cite{TaoLi_116Sn,PATEL2012447,KBHOWARD_moly,npa-garg,JARROYO2025, Howard2019,PATEL2013178,Yogesh90Zr} by utilizing $\alpha$-particle inelastic scattering in finite nuclei to provide necessary ISGMR data for the isotopic chain $^{104,106,108,110}$Pd, in an attempt to study the range of even-even isotopes between \nuc{90}{Zr} and tin isotopes where softness was first observed. \textbf{Section II} touches briefly on experimental details of this study. We also discuss the optical model parameters for $\alpha$-particle scattering off $^{104,106,108}$Pd, arrived at via $\chi^2$ fitting of the elastic cross sections.

Further, we describe the multipole decomposition analysis (MDA) procedure, including details of the affine-invariant Markov chain Monte Carlo (MCMC) method employed in this analysis. \textbf{Section III} shows the results of the MDA; the giant resonance strength distributions for $^{104,106,108,110}$Pd extracted therefrom; and the Lorentzian parameters from fits to the strength distributions. Also provided in this section are the energies derived from moment ratios of the strength distributions. \textbf{Section IV} provides a comparison of the experimental results with calculations utilizing self-consistent QRPA plus Quasiparticle Vibration Coupling (QPVC) effects. \textbf{Section V} provides a summary of the results within the context of the softness conundrum and the current discourse regarding the state of models for bulk nuclear matter.

\section*{\label{methods}II. Experimental Methods}\label{II}

\begin{figure*}[]
    \centering
    \includegraphics[width=1.0\textwidth]{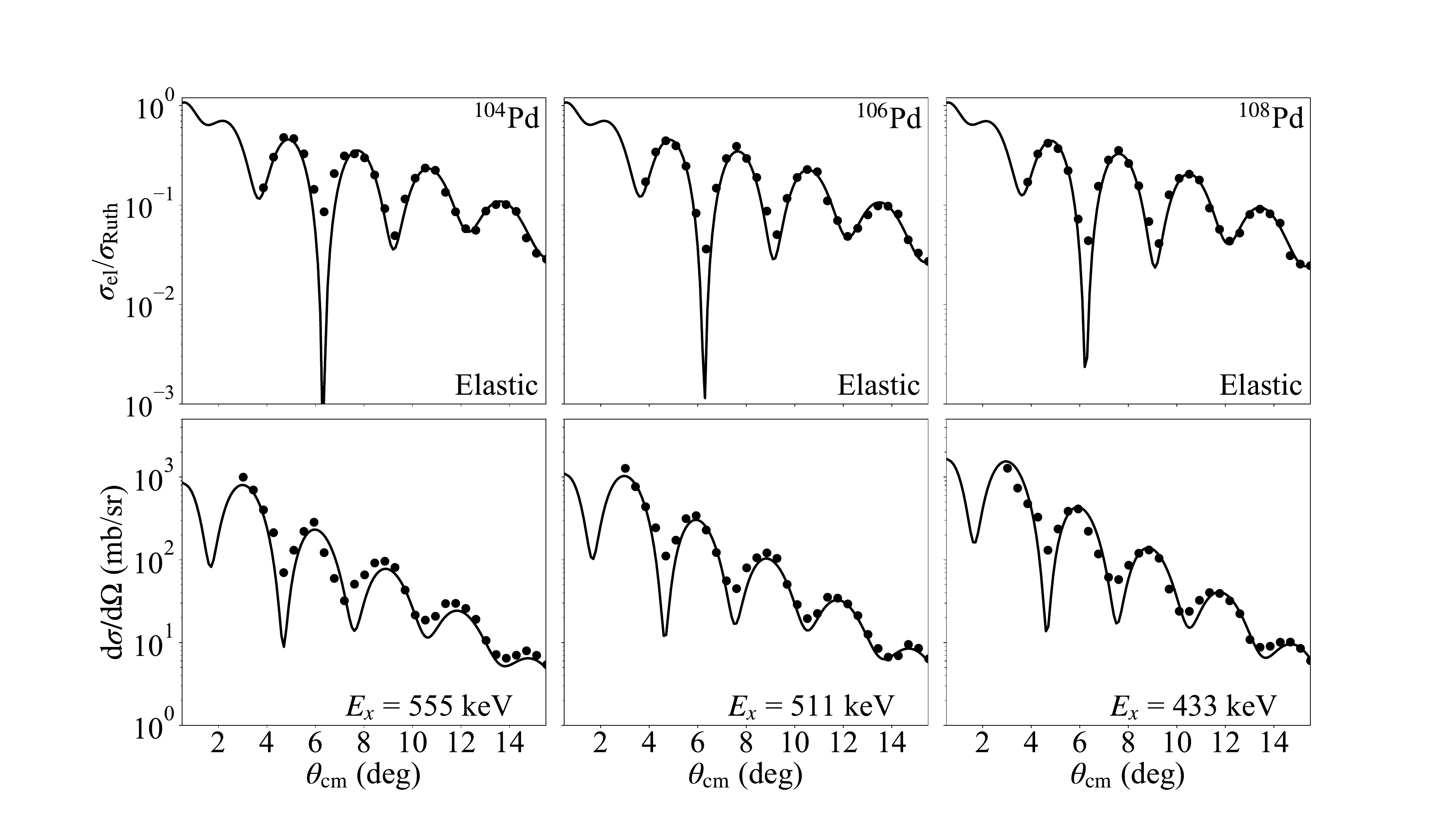}

    \caption{Experimental data (filled circles) for the elastic scattering cross sections normalized to the Rutherford cross sections (top panels) and the low-lying $2_1^+$ states (bottom panels) are shown. Optical model fits (solid lines) to elastic scattering data for all nuclei under investigation in this work are shown in the top panels. DWBA calculations (solid lines) for the corresponding low-lying states were carried out using the extracted OMP values and the corresponding adopted $B(E\mathrm{2})$ values from \cite{KIBEDI200235,RAMAN20011} }
    \label{fig:OMPs}
\end{figure*}

 The facilities at RCNP, the University of Osaka, have been well documented in previous publications on ISGMR \cite{JARROYO2025, Howard2019,PATEL2013178,Yogesh90Zr}, and will not be covered in great detail here. As well, a detailed description of Grand Raiden spectrometer can be found in Refs. \cite{GRaiden,Fujiwara2023}. A halo-free $\alpha$-particle beam of  386\,-MeV energy impinged on self-supporting targets of $^{104,106,108,110}$Pd. Table~\ref{tab:table1} shows the areal densities and isotopic enrichment of each target. Target thicknesses were obtained from energy-loss and stopping-range measurements with $\alpha$ particles from $^{148}$Gd, $^{241}$Am, $^{244}$Cm sources.  The Grand Raiden spectrometer was employed to obtain elastic and inelastic scattering spectra for all targets. Focal plane detectors enabled precise momentum measurement of scattered $\alpha$-particles. The MWDC were followed by two plastic scintillators that provided particle identification via measurement of the time-of-flight and energy-loss correlations. 

\setlength{\tabcolsep}{14pt} 
\renewcommand{\arraystretch}{1.2} 
\begin{table}[b]
\caption{
\label{tab:table1}
Areal densities and enrichment for all targets investigated in this work.
}
\begin{ruledtabular}
\begin{tabular}{lcdc}

\textrm{Target}&\textrm{Areal Density } & \multicolumn{1}{c}{Enrichment}\\
\colrule
\nuc{104}{Pd} & 5.10\,mg/cm$^2$ & 95.0 \%\\
\nuc{106}{Pd} & 3.85\,mg/cm$^2$ & 96.7 \%\\
\nuc{108}{Pd}& 5.16\,mg/cm$^2$ & 98.5 \%\\
\nuc{110}{Pd}& 3.88\,mg/cm$^2$ & 97.7 \%\\
\end{tabular}
\end{ruledtabular}
\end{table}

Inelastic scattering data were obtained in the angular range of 0.0\degree< $\theta_{\mathrm{lab}}<$ 9.2\degree in order to best observe the ISGMR due to its forward peaked angular distribution. Spectrographic averaging results were determined in the true angular range of  $\sim$0.7\degree $< \theta_{\mathrm{avg}}<$ 9.0\degree (see Ref. \cite{Howard2019} for details on spectrographic averaging), depending on the binning of the 0\degree data. Inelastically-scattered $\alpha$ particles were measured over an excitation energy of 8--35\,MeV. Examples of forward-angle spectra are shown in Fig.~\ref{fig:fwd_spectra}.

\setlength{\tabcolsep}{6pt}
\renewcommand{\arraystretch}{1.1}
\begin{table*}[t]
\caption{Fermi density parameters from Ref. \cite{FRICKE1995177}, real potential depth, and Woods-Saxon imaginary potential parameters for the $^{104,106,108}$Pd nuclei. Nuclear half-mass radii, $c$, and diffuseness, $a$, were fixed parameters and not varied during the OMP searches. $V_R$ is the depth of the single-folded potential from equation (\ref{eq:rvol}), $W_{I}$, $r_{0,I}$, and $a_{I}$ are, the depth, reduced radius, and diffuseness of the imaginary volume potential from equation (\ref{eq:wvol}), respectively. The $B(E2)$ values and the excitation energies of the 2$_1^+$ are provided in the two right-most columns.
\label{tab:table2}}
\begin{ruledtabular}
\begin{tabular}{cccccccccc}
 \multirow{2}{*}{Nucleus}&\multicolumn{2}{c}{Density Parameters}&\multicolumn{4}{c}{Optical Model Parameters}&\multicolumn{2}{c}{2$_1^+$}\\ \cline{2-9}
 &c    & $a$ & $V_R$ & $W_I$ & $r_{0,I}$ & $a_I$ & $B(E2)$ & $E_{x}$ \\
 &(fm) &(fm) &(MeV)  & (MeV) &           & (fm)  & ($e^2b^2$)    & (keV)\\
 \cline{1-9}
 \\
 \nuc{104}{Pd} & 5.25 & 0.525 & 36.2 & 33.7 & 1.03 & 0.72 & 0.535 & 555 \\ \\
 \nuc{106}{Pd} & 5.28 & 0.525 & 37.5 & 31.0 & 1.05 & 0.71 & 0.660 & 511 \\ \\
 \nuc{108}{Pd} & 5.32 & 0.525 & 36.4 & 32.1 & 1.05 & 0.78 & 0.760 & 439 \\ \\
  \nuc{110}{Pd} & 5.35 & 0.525 & -- & -- & -- & -- & -- & -- \\
\end{tabular}
\end{ruledtabular}
\end{table*}

Elastic scattering measurements were performed over an angular range of 2\degree $<\theta_\mathrm{{lab}}<$ 16\degree. Data were taken for $^{104,106,108}$Pd (elastic scattering data could not be obtained for $^{110}$Pd due to beam-time constraints). As optical model parameters (OMPs) vary smoothly over mass numbers, OMPs for $^{108}$Pd were used for analysis of the $^{110}$Pd inelastic scattering spectra. Field strengths of the Grand Raiden magnets for the elastic scattering runs were tuned for a focal-plane acceptance energy range of 0--10\,MeV, enabling observation of the low-lying excited states. Calibration runs with a $^{24}$Mg target were taken intermittently throughout target run time at central angles corresponding to the current loaded target to calibrate the energy of the inelastic scattering spectra. Various excited states were used throughout this process by comparing calibration spectra with previously collected high energy-resolution \nuc{24}{Mg}($\alpha$,$\alpha^\prime$)\nuc{24}{Mg}$^*$ data at RCNP at a wide range of angles ejectile angles. Prominent excited states at a particular ejectile angle were then used to energy calibrate target spectra.

\begin{figure*}[]
    \includegraphics[width=1.\textwidth, height=5.cm]{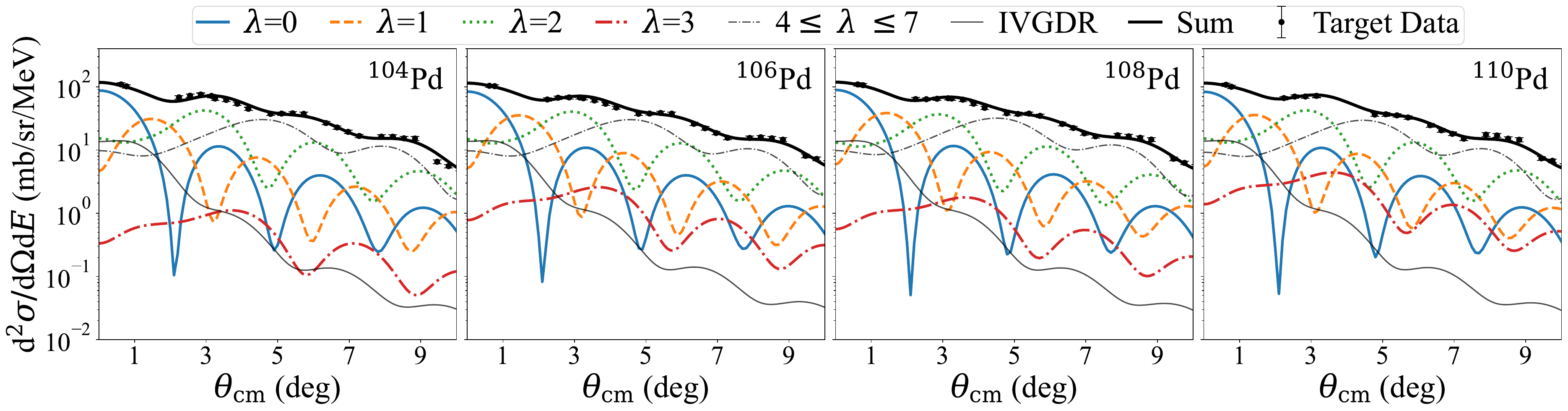}
    \caption{
    (Color online) Results of MDA for all targets for an excitation-energy bin centered at 16.5\,MeV. In each case, the filled black circles are the experimental cross sections and the solid black line is the sum of all multipolarities included in the MDA up to $\lambda_{max}=7$. The blue line is the ISGMR response. The dashed orange, dotted green, and dash-dotted red are the ISGDR, ISGQR, and $L$=3 response, respectively. The grey dash-dotted line is the sum of the higher order responses, i.e. $\lambda=4$ to $\lambda_{max}=7$, and the solid grey line is the IVGDR response that was fixed for the MDA based on photonuclear cross sections.
    }
    \label{fig:dwba}
\end{figure*}

Analysis of the spectra in this work, performed in ROOT~\cite{root},  differs minimally from previous efforts~\cite{JARROYO2025, Howard2019}, and is only briefly summarized here. Energy losses in the scintillator spectra enabled particle identification gating to only $\alpha$-ejectiles. Grand Raiden's double-focusing mode was used for off-median instrumental background subtraction; this technique is described in a recent publication \cite{JARROYO2025}. Inelastic scattering spectra are sliced by energy and angle, with 0\degree runs spanning $\pm$ 0.6\degree and the others spanning $\pm$ 0.8\degree. Forward-angle runs were split into 3 bins, with outer bins merged to both improve statistics in the off-center regions and to assist in disentangling the effects of other giant resonances from the ISGMR. Upstream analysis also entailed observing the subtraction spectra, where the 1.4\degree data are subtracted from those at $\theta_\mathrm{avg}=0.6$\degree. The subtraction spectra exhibit primarily the ISGMR strength, due to the forward peaked nature of the ISGMR and relatively flat responses of other giant resonances in this angular regime \cite{BRANDENBURG198729}. Fig.~\ref{fig:fwd_spectra} shows the original forward angle spectrum, as well as the spectrum following the neighboring 1.4\degree spectrum subtraction to demonstrate this.

Optical model parameters (OMP) were obtained from the measured elastic scattering angular distributions using the $\chi^2$ fitting routine in the code PTOLEMY~\cite{macfarlane_ptolemy}. The optical model approach for $\alpha$-inelastic scattering is well established \cite{KhoaDF, UGarg}, and this work does not deviate from the formalism so as to keep results standardized with previous efforts from ISGMR studies at RCNP. This approach utilizes a hybrid density-dependent, single-folded potential, with a Gaussian interaction kernel in combination with a Woods-Saxon absorptive term and a Coulomb term: 
\begin{equation}
    U(r) = V_{\mathrm{DDG}}(r) + iW_{\mathrm{Vol}}(r) + V_{\mathrm{Coul}}(r),
\end{equation}
where the Coulomb potential is the standard interaction between a point and a uniform sphere. The absorptive volume potential has the standard form factor:
\begin{equation} \label{eq:wvol}
    \begin{aligned}
        W_{\mathrm{Vol}}(r) & = W_Vf(r,R_V,a_V,) \\
        f(r,R_{V},a_{V}) & = \left[1+\exp{\frac{r-R_{V}}{a_{V}}}\right]^{-1}.
    \end{aligned}  
\end{equation}

For the real volume potential, Satchler and Khoa have demonstrated that a single-folded, density-dependent potential is sufficient for characterizing this term for ($\alpha$,~$\alpha^\prime$) reactions \cite{SatchlerSF}, so that:

\begin{equation} \label{eq:rvol}
    V_{\mathrm{DDG}}(r) = V_V \int \rho(r')F(\rho)v_G(s) \mathrm{d}^3r'.   
\end{equation}

The single-folding approach uses a simplified, parameterized Gaussian interaction which is folded over the target-density distribution $\rho(r')$, with 

\begin{align}
    F(\rho) &= 1 - \zeta \rho^\beta(r'), & v_G(s) &= V_R \exp(-|\mathbf{r}-\mathbf{r}'|/t^2); \nonumber
\end{align}

\noindent the numerical parameters from Ref.~\cite{SatchlerSF} are $\zeta~\approx~1.9\ $fm$^2$, $\beta=2/3$, and $t\approx 1.88\ $fm. These are empirically tested parameters created in an attempt to go beyond the Woods-Saxon deformed models.

Parameter sweeps of the optical model potentials were carried out using \nuc{110}{Cd} parameters from Ref. \cite{PATEL2012447} as the starting point for each set of potentials until the reduced $\chi^2$ values were stationary. Density parameters remain fixed, and were obtained from Ref. \cite{PhysRevC.63.034007}; these are listed in Table \ref{tab:table2}. Figure \ref{fig:OMPs} shows fits to the elastic scattering angular distributions for all Pd isotopes studied in this work. The resulting OMPs were verified by comparing the experimental angular distributions for the low-lying 2$_1^+$ states with those calculated with these OMPs using the accepted values for the $B(E\mathrm{2})$ transition probabilities \cite{KIBEDI200235,RAMAN20011}. As shown  in Fig. \ref{fig:OMPs}, the agreement between the experimental and calculated angular distributions is very good in all cases, establishing the validity of the extracted OMPs.

\begin{figure}[h]

    \includegraphics[width=0.4\textwidth]{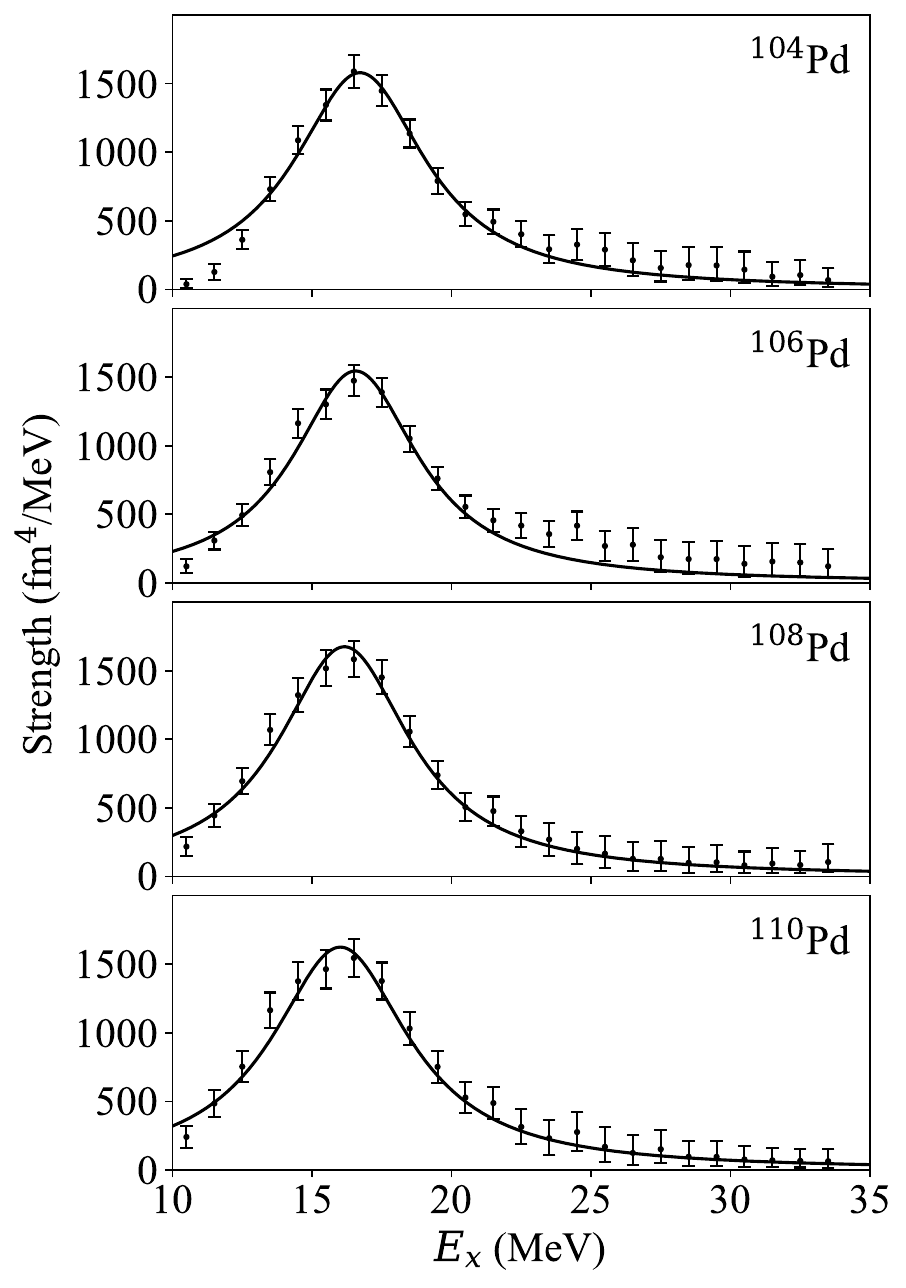}
    \caption{
    Extracted ISGMR strength distributions (filled black circles with error bars) for all nuclei investigated in this work. Also shown superimposed are  Lorentzian fits (solid lines) to the data.
    }
    
    \label{fig:ISGMR_strs}
\end{figure}

Distorted Wave Born Approximation (DWBA) calculations were performed using PTOLEMY to
determine the angular distribution for each of the electric giant resonance to be included in the MDA for 100\% of the energy-weighted sum rule (EWSR). This was done in steps of  1-MeV excitation energy over 8--35\,MeV. Density deformation parameters were calculated using the form factors from Refs.~\cite{Harakeh, Howard:PhD:2020}:

\begin{equation}
    \begin{aligned}
    \delta\rho_{(\lambda=0)} &= -\frac{\beta_0}{r^2} \frac{\mathrm{d}}{\mathrm{d}r} (r^3\rho(r))  \\
    \delta \rho_{(\lambda = 1)} &= -\frac{\beta_1}{c}\left[ 3r^2\frac{\mathrm{d}^2}{\mathrm{d}r^2} +10r - \frac{5}{3}\langle r^2 \rangle \frac{\mathrm{d}}{\mathrm{d}r} + \epsilon \left( r \frac{\mathrm{d}^2}{\mathrm{d}r^2} +4 \frac{\mathrm{d}}{\mathrm{d}r} \right) \right]  \rho(r) \\
    \delta \rho_{(\lambda>1)} &= -\beta_\lambda c \frac{\mathrm{d}\rho(r)}{\mathrm{d}r}.
    \end{aligned}
\end{equation}

Transition potential form factors are approximated using the respective density deformation parameter equation, calculated from the ground-state potential. Single folding was calculated using an in-house code bench-marked against SDOLF/DOLF \cite{Howard:PhD:2020}.

The calculated angular distributions are input in the MDA. Following previous work \cite{JARROYO2025, Howard2019}, this utilized affine invariant unsupervised learning algorithms of Ref.~\cite{2013PASP..125..306F}. Inelastic angular distributions for each energy bin are fitted to a sum of the DWBA cross sections corresponding to 100\% energy-weighted sum rule (EWSR) for the respective multipolarities:
\begin{small}
\begin{equation}
    \hspace{-0.2cm} \frac{\mathrm{d}^2\sigma^\mathrm{exp}(\theta_\mathrm{cm},E_{x})}{\mathrm{d}\Omega \mathrm{d}E} = \sum_{\lambda = 0}^{\mathrm{\lambda_{max}}}a_{\mathrm{\lambda}}(E_{x}) \frac{\mathrm{d}^2\sigma^\mathrm{DWBA}_\mathrm{\lambda}(\theta_\mathrm{cm},E_{x})}{\mathrm{d}\Omega \mathrm{d}E}. \label{eq:isgqr}
\end{equation}
\end{small}

\noindent The affine-invariance assists in addressing multicollinearities when fitting the continuum space that may arise due to the similar angular character of both even and odd giant resonances after their first maximum. The MCMC procedure maximizes the likelihood distribution of the $\chi^2$ with affine-invariant steps in the $\lambda_\mathrm{max}$-dimensional parameter space, increasing convergence speed compared to a more traditional sampling routine with highly correlated parameters. More explicit details on this fitting approach can be found in Ref. \cite{2013PASP..125..306F}. 

\begin{table*}[]
\caption{\label{tab:table3}
ISGMR parameters for the Pd isotopes extracted in this work. $E_\mathrm{centroid}$ is the centroid energy resulting from each Lorentzian fit and $\Gamma$ is the width from (\ref{eq:lorentzian}). Moment ratios and EWSRs are calculated from the Lorentzians themselves over the full energy range of 8--30\,MeV. The uncertainties listed for the EWSRs are a combination of statistical effects from the MCMC sampling as well as uncertainties from cross section calculations and do not include systematic uncertainties of 15\%--20\% arising from the DWBA calculations and the choice of OMPs.
}
\begin{ruledtabular}
\setlength{\tabcolsep}{4pt}
\begin{tabular}{ccccccr}
 \multirow{2}{*}{Nucleus}    & $E_{\mathrm{centroid}} $ & $\Gamma$ & $\frac{m_1}{m_0}$ & $\sqrt{\frac{m_1}{m_{-1}}}$ & $\sqrt{\frac{m_3}{m_1}}$ & EWSR \\
   & (MeV) & (MeV) & (MeV) & (MeV) & (MeV) & \%\\  \hline 
\\
\nuc{104}{Pd} & 16.7$\pm$0.1 &2.9$\pm$0.2& 16.7$\pm$0.1 & 16.3$\pm$0.1  & 17.0$\pm$0.1  & 114$^{+4}_{-5}$ \\ \\
\nuc{106}{Pd} & 16.6$\pm$0.1 &2.7$\pm$0.2& 16.5$\pm$0.1 & 16.2$\pm$0.1  & 16.9$\pm$0.1  & 105$^{+4}_{-5}$ \\ \\
\nuc{108}{Pd} & 16.2$\pm$0.1 & 2.9$\pm$0.2 & 16.2$\pm$0.1  & 15.9$\pm$0.1 & 16.6$\pm$0.1 & 111$^{+3}_{-4}$ \\ \\
\nuc{110}{Pd} & 16.0$\pm$0.2 & 3.0$\pm$0.2 & 16.1$\pm$0.1  & 15.7$\pm$0.1 & 16.5$\pm$0.2 & 108$^{+2}_{-5}$ \\
 
\end{tabular}
\end{ruledtabular}
\end{table*}

Multipolarities of up to $\lambda = 7$ were included to model the nuclear continuum, as knock-out/exchange reactions mimicking pure multipoles may be present.  MDA results did not change in any appreciable manner by including $\lambda > 7$. In this work, it has been possible to reliably extract the strength distributions only for the ISGMR, the isoscalar giant dipole resonance (ISGDR) and the ISGQR in all the nuclei under investigation. However, inclusion of higher order terms assists with the analysis by absorbing non-resonance responses that a MCMC fitting routine would otherwise misattribute to the ISGDR or ISGQR. 

The contribution of the isovector giant dipole resonance~(IVGDR) was not a free parameter in these fits. Those angular distributions were calculated using the adopted strength parameters from photonuclear cross sections \cite{IVGDR}. The ratio of the strengths between the energy bin and total strength was used to calculate IVGDR strength percentage, $a_\mathrm{IVGDR}$, and the angular distributions calculated for it via PTOLEMY. Sample MDA results are shown in Fig. \ref{fig:dwba}.

\section*{\label{sec:results}III. Results}\label{III}

The strengths for various multipoles are calculated using the scaling parameters from these sampling routines, $a_\lambda$, and the corresponding 100\% EWSR strengths. For the monopole response, this takes the form:  

\begin{equation}
    S_0(E_{x}) = \frac{2 \hbar^2 A \langle r^2 \rangle}{m E_{x}} a_0(E_{x}), \label{eq:isgmr}
\end{equation}

\noindent where $m$ is the nucleon mass, $A$ is the mass number of the target, and  $\langle r^2 \rangle$ is the mean-square radius of the ground-state density distribution calculated using a Woods-Saxon 
form~\cite{UGarg,Harakeh}. For other giant resonances, the strength can similarly be calculated using their form factors from Ref. \cite{Harakeh}. For $\lambda=1$ and $\lambda = 2$:

\begin{equation}
    S_1(E_{x}) =\frac{3\hbar^2A}{32\pi mE_x}\bigg(11\langle r^4\rangle - \frac{25}{3} \langle r^2\rangle^2 - 10\epsilon \langle r^2\rangle \bigg) a_1(E_{x}) \label{eq:isgdr}
\end{equation}
and
\begin{equation}
    S_2(E_{x}) =\frac{50\hbar^2A}{8\pi mE_x} \langle r^2\rangle a_2(E_{x}). \label{eq:isgqr}
\end{equation}

\subsection{ISGMR Response}

Useful comparison points for theory and experiment are energies derived from moment ratios of the strength distributions. For the monopole resonance, there are three calculable quantities of interest:
\begin{equation}
    \begin{aligned}
        E_\mathrm{constrained} & = \sqrt{\frac{m_1}{m_{-1}}} \\
        E_\mathrm{centroid} & = \frac{m_1}{m_0} \\
        E_\mathrm{scaling} & = \sqrt{\frac{m_3}{m_1}},
    \end{aligned}
\end{equation}
with $m_k = \int \mathrm{d}E_{x}S_0(E_{x})E_{x}^k.$

Additionally, ISGMR strength distributions for medium-heavy nuclei are often fit by Lorentzians to interpret results independent of potentially spurious strength outside of main response regime of the ISGMR. The fit function adopted for this analysis is:
\begin{equation}
L(E_{x}, I_0, E_{\mathrm{center}}, \Gamma) =I_0\frac{\Gamma}{(E_{x}-E_{\mathrm{center}})^2 + \Gamma^2}. \label{eq:lorentzian}
\end{equation}

\begin{figure}[b]

    \includegraphics[width=0.4\textwidth]{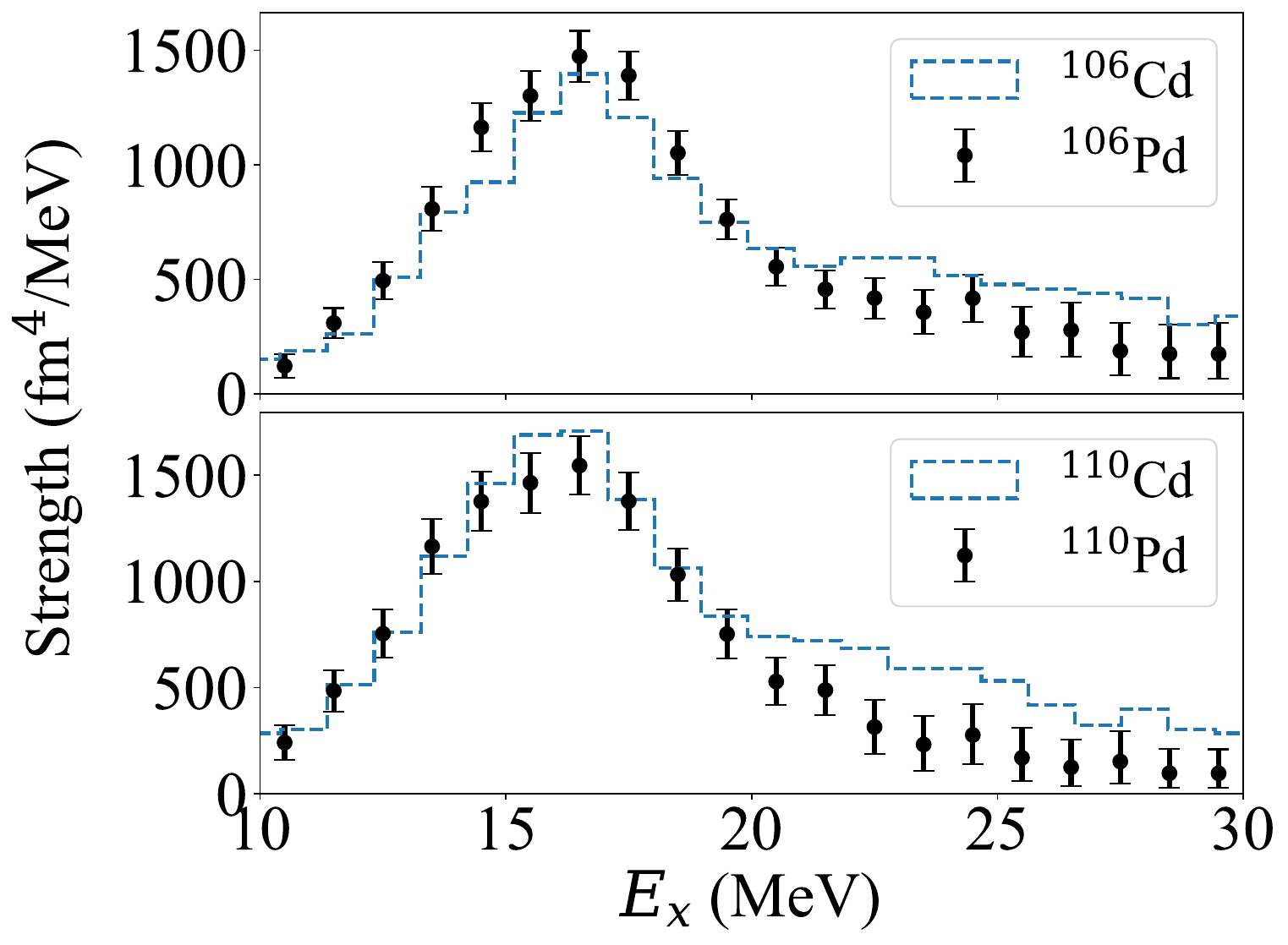}
    \caption{
    (Color online) Comparison between extracted ISGMR strengths for \nuc{106,110}{Pd} and \nuc{106,110}{Cd}. Filled black circles with error bars are \nuc{106}{Pd} (top panel) and \nuc{110}{Pd} (bottom panel), whereas the dashed histograms are for the ISGMR strengths of \nuc{106}{Cd} and \nuc{110}{Cd} obtained from Ref. \cite{PATEL2012447}. This isobaric comparison was done as a benchmark to assess the efficacy of extractions in this work.
    }
    \label{fig:ISGMR_bm}
\end{figure}

Table \ref{tab:table3} shows fit parameters for ISGMR strength distributions of \nuc{104,106,108,110}{Pd}, as well as $E_\mathrm{centroid}$, $E_\mathrm{constrained}$, and $E_\mathrm{scaling}$, and the total exhausted EWSR. Figure \ref{fig:ISGMR_strs} shows the raw strength distribution from the MCMC fit routine from Eq.~(\ref{eq:isgmr}), as well as the Lorentzian fits utilizing Eq.~(\ref{eq:lorentzian}). It should be noted that the uncertainties in EWSR shown here, and in subsequent figures showing extracted strengths for various multipoles, are statistical from the MDA process combined with the experimental errors from cross-section extraction. The moment ratios presented here were calculated from the Lorentzian fits themselves to best predict the constrained, centroid, and scaling energies. The EWSRs were also calculated with the Lorentzian fits, rather than a direct sum of the raw distributions. This was done in order to avoid ambiguities with choice of excitation energy range for determining these values. If a direct sum of the raw distributions is utilized instead of Lorentzian fits, an excitation energy range of 10--20\,MeV results in EWSRs of 80--90\%

ISGMR responses among all Pd isotopes studied here exhibit the nominal properties. For example, there is the expected inverse dependence with $A$ of the ISGMR energy, both from Lorentzian fits and from moment ratios, for all studied targets.

Figure \ref{fig:ISGMR_bm} shows a comparison of the ISGMR strength distributions in the two cases of Pd-Cd isobars as benchmarks for the response from this analysis. As is evident from these figures, the present results agree very well with the results from Patel et al. \cite{PATEL2012447}. Further, the reported centroids for the ISGMR strengths in \nuc{106}{Cd}  (16.5$\pm$0.2\,MeV) and \nuc{110}{Cd} (16.1$\pm$0.2\,MeV),  are fully consistent with results presented here. This isobaric comparison further affirms the validity of the analysis performed here.

Analyses combining ISGMR studies of isotopic chains with the leptodermous expansion of $K_A$ can lead to various parameters associated with nuclear incompressibility \cite{TaoLi_116Sn, PATEL2012447}. In the leptodermous expansion, the finite incompressibilities $K_A$ can be expanded as \cite{BLAIZOT1980171}:

\begin{equation}
    K_{A} \approx K_\infty + K_\mathrm{surf}A^{-1/3} + K_\tau\eta^2 +  K_\mathrm{Coul}Z^2A^{-4/3},
    \label{eqn:leoptodermous_expansion}
\end{equation}

\noindent with $K_\mathrm{surf}$, $K_\tau$, and $K_\mathrm{Coul}$ corresponding to the surface term, symmetry term, and the Coulomb term from the liquid drop model, $\eta= \frac{N-Z}{A}$ is the asymmetry term. Equation (\ref{eqn:leoptodermous_expansion}) can be rearranged, isolating known terms and noting that $K_\infty/K_\mathrm{surf}\sim -$1 \cite{UGarg} and that $A^{-1/3}$ varies slowly over isotopic chains of medium-heavy nuclei, leading to a straightforward relationship between $K_A$, $K_\mathrm{Coul}$, and $K_\tau$ as a function of $\eta^2$:

\begin{equation}
    K_A - K_\mathrm{Coul}Z^2A^{-4/3} \approx \mathrm{Constant} + K_\tau\eta^2.
    \label{eqn:ktau_finder}
\end{equation}

\begin{figure}
    \centering
    \includegraphics[width=0.5\textwidth]{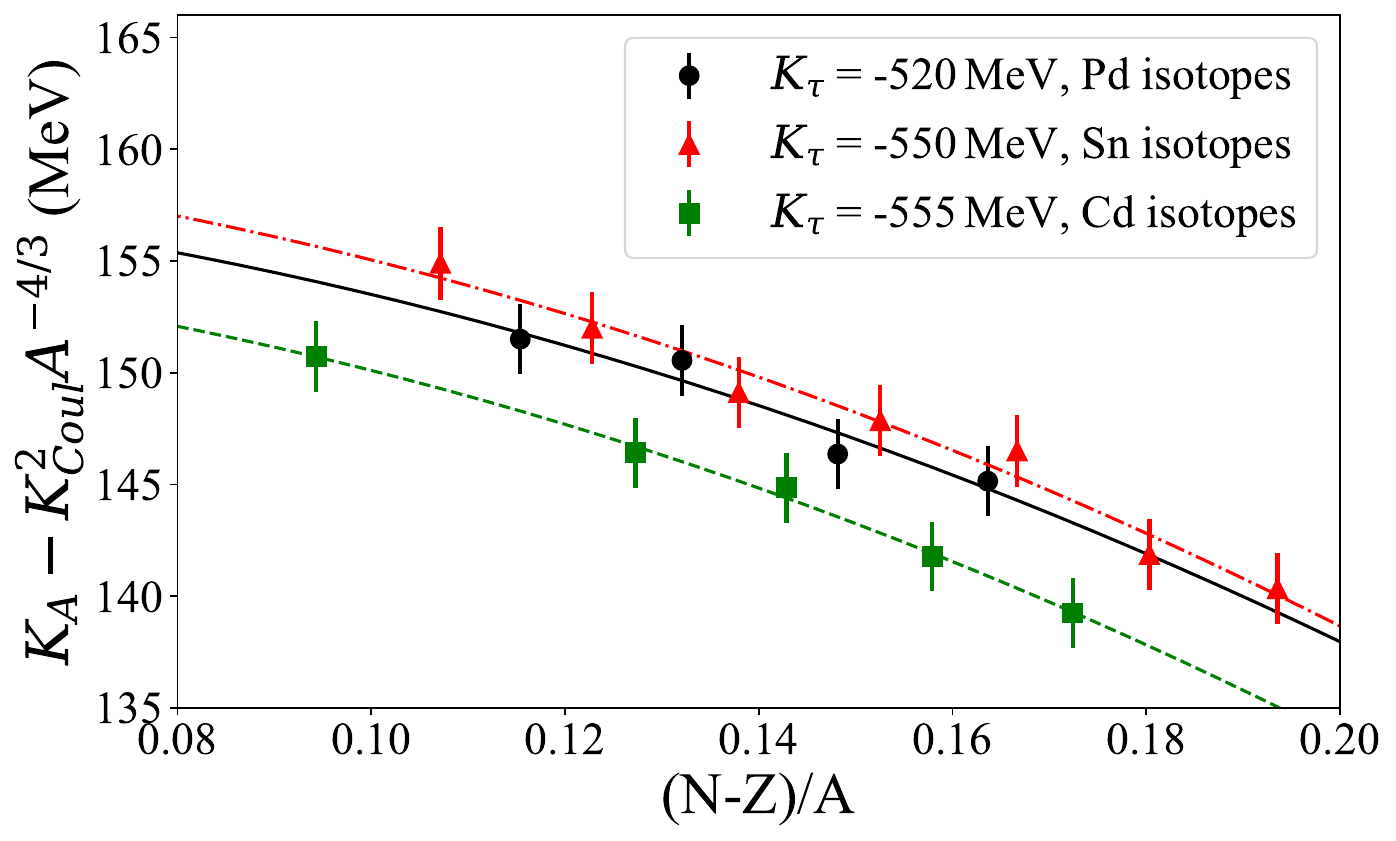}
    \caption{
    (Color online) Quadratic fits using Eq.~(\ref{eqn:ktau_finder}) to arrive at values of $K_\tau$ via the leptodermous expansion. The results from the present work on the Pd isotopes (black line, circles) is compared to that of Refs. \cite{TaoLi_116Sn} (red dash-dotted line, triangles) and \cite{PATEL2012447} (green dashed line, squares) for the Sn and Cd isotopes, respectively.
    }
    \label{fig:ktau_calculation}
\end{figure}

Equation (\ref{eqn:ktau_finder}) was used to extract a value for the asymmetry term of nuclear incompressibility, $K_\tau$, which may provide insights into the incompressibility of asymmetric nuclear matter~\cite{UGarg}. Finite compressibilities $K_A$ were calculated using the constrained energy from the extracted strength distribution for direct comparison with previous works, using:

\begin{equation}
    E_\mathrm{ISGMR} = \hbar\sqrt{\frac{K_A}{m\langle r_0^2\rangle}}.
\end{equation}

\noindent Figure \ref{fig:ktau_calculation} shows the results of this analysis for the Pd isotopes studied in this work; also presented for comparison purposes are the results from the previous work on the Cd and Sn isotopes \cite{TaoLi_116Sn, PATEL2012447}. This analysis arrived at a value of $K_\tau=-520\pm140$\ MeV. A quadratic fit to calculated $K_A$s resulted in $\pm$100\,MeV of error attributed purely to statistics. The remaining $\pm40$\,MeV is attributed to uncertainty in $K_\mathrm{Coul}$ itself. Despite larger errors due to a large, constrained energy gap between \nuc{106}{Pd} and \nuc{108}{Pd}, this value of $K_\tau$ is still fully consistent with the previous results.

\begin{figure}
    \includegraphics[width=0.4\textwidth]{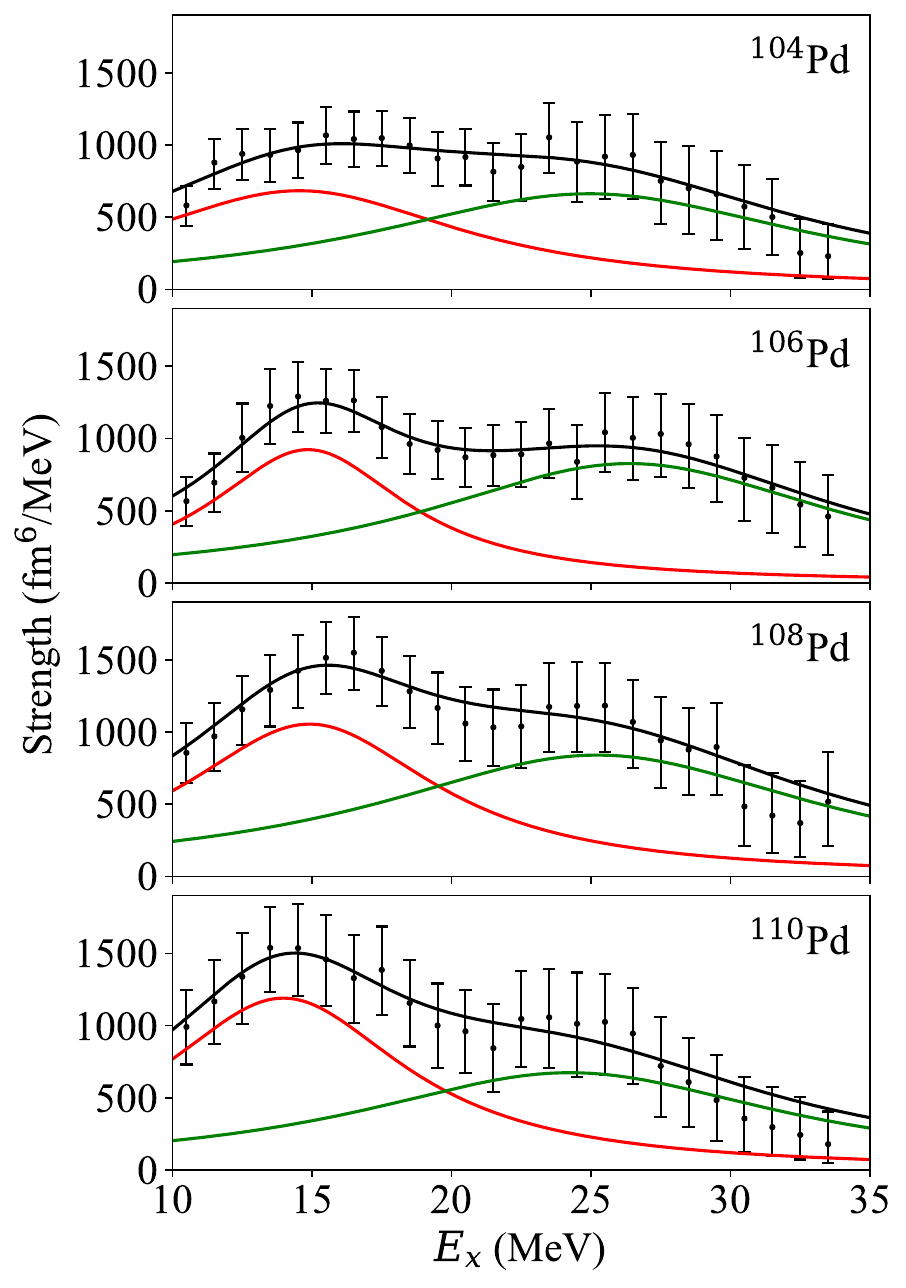}
    \caption{(Color online) Extracted ISGDR strength distributions (filled circles with error bars) for all nuclei investigated in this work. Also shown superimposed are  two-Lorentzian fits to the data. The red and green solid lines are the LE and HE fits to the strength distribution, respectively, and the solid black lines are the sum of both constituent Lorentzians.
    } 
    \label{fig:ISGDRs}
\end{figure}

\begin{table*}[]
\renewcommand{\arraystretch}{1.1}
\caption{
LE and HE fit parameters for the ISGDR obtained in this work. $E_\mathrm{LE(HE)}$ are the low (high) energy centroids from Lorentzian fits and $\Gamma_\mathrm{LE(HE)}$ are the low (high) energy widths. The uncertainties listed for the EWSRs are a combination of statistical effects from the MCMC sampling as well as uncertainties from cross section calculations and do not include systematic uncertainties of 15\%--20\% arising from the DWBA calculations and the choice of OMPs.
\label{tab:table4}
}
\begin{ruledtabular}
\setlength{\tabcolsep}{2pt}
\begin{tabular}{lllrllr}
 \multirow{2}{*}{Nucleus} &\multicolumn{3}{c}{LE Parameters}&\multicolumn{3}{c}{HE Parameters} \\ \cline{2-4} \cline{5-7}
 &$E_{\mathrm{LE}} $ & $\Gamma_{\mathrm{LE}}$ & EWSR & $E_{\mathrm{HE}} $ & $\Gamma_{\mathrm{HE}}$ & EWSR \\ 
 &(MeV) &(MeV) & \% & (MeV) & (MeV) & \%\\
 \cline{1-7}
 \\
 \nuc{104}{Pd} & 14.5$^{+1.0}_{-1.3}$ & 7.2$^{+1.8}_{-2.0}$ & 43$^{+13}_{-15}$& 24.9$\pm$1.8 & 9.5$^{+1.7}_{-2.7}$ &  59$^{+8}_{-10}$ \\ \\
 \nuc{106}{Pd} & 14.8$^{+0.6}_{-0.7}$ & 4.3$^{+1.9}_{-1.3}$ & 41$^{+13}_{-12}$& 26.4$^{+1.1}_{-1.6}$ &  9.1$^{+2.0}_{-2.8}$ & 67$^{+6}_{-7}$ \\ \\
 \nuc{108}{Pd} & 14.9$^{+0.6}_{-0.9}$ & 5.5$^{+1.7}_{-1.4}$ & 54$^{+9}_{-14}$ & 25.2$^{+1.7}_{-1.9}$ & 9.7$^{+1.6}_{-2.5}$ &69$^{+5}_{-6}$ \\ \\
 \nuc{110}{Pd} & 14.0$^{+0.8}_{-1.0}$ & 5.4$^{+1.4}_{-1.3}$ & 54$^{+8}_{-15}$ & 24.2$^{+1.9}_{-1.5}$ & 9.3$^{+1.9}_{-2.9}$ &54$^{+6}_{-7}$ \\
\end{tabular}
\end{ruledtabular}
\end{table*}

\subsection{ISGDR Response}

Figure \ref{fig:ISGDRs} shows the extracted ISGDR strength distributions for the nuclei investigated in this work. As has been seen previously in other nuclei \cite{UGarg}, this strength has two distinct components, referred to generally as the low-energy (LE) component and the high-energy (HE) component. The strength distributions were, therefore, fitted with two Lorentzians; Table \ref{tab:table4} lists the parameters of the Lorentzian fits as well as the EWSRs that correspond to each Lorentzian. The listed EWSRs were calculated from the Lorentzian fits themselves separately for the LE and HE components. Total EWSRs from summing LE and HE components range between $\sim$80--120\%. The LE centroid is not dependent on $K_\infty$, whereas the HE component is sensitive to this parameter in the equation of state. The fact that the strength distributions for the ISGDR are well constrained in the high-energy regime, lends confidence to the location of the HE-ISGDR.

\subsection{ISGQR Response}

\begin{figure}[b]
    \includegraphics[width=0.4\textwidth]{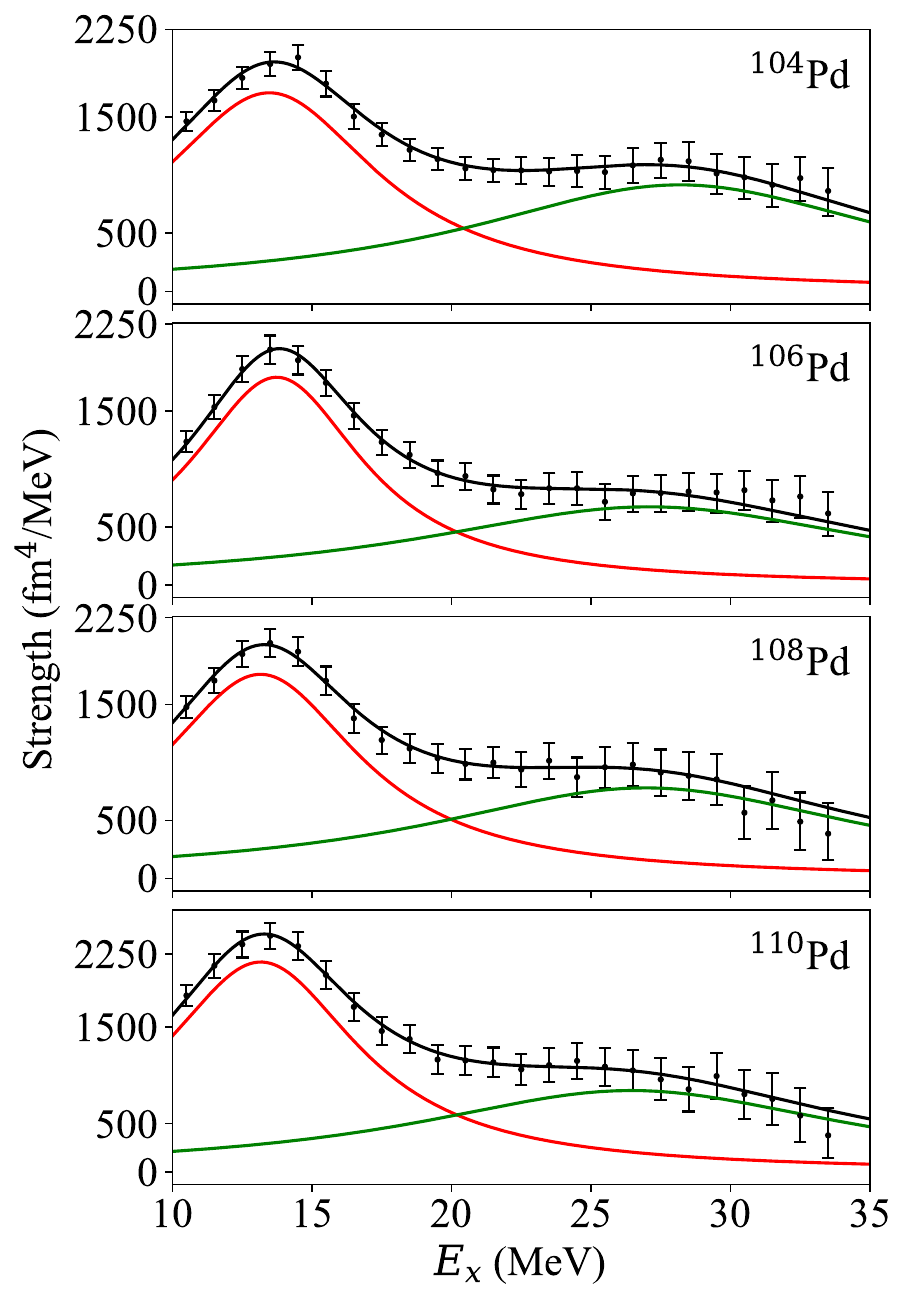}
\caption{(Color online) Extracted ISGQR strength distributions (filled circles with error bars) for all nuclei investigated in this work. Also shown superimposed are  two-Lorentzian fits to the data. The red and green solid lines are the LE and HE fits to the strength distribution, respectively, and the solid black lines are the sum of both constituent Lorentzians.
    } 
    \label{fig:ISGQRs}
\end{figure}

\begin{table*}[]
\renewcommand{\arraystretch}{1.1}
\caption{
LE and HE fit parameters for the ISGQR obtained in this work. $E_\mathrm{LE(HE)}$ are the low (high) energy centroids from Lorentzian fits and $\Gamma_\mathrm{LE(HE)}$ are the low (high) energy widths. The uncertainties listed for the EWSRs are a combination of statistical effects from the MCMC sampling as well as uncertainties from cross section calculations and do not include systematic uncertainties of 15\%--20\% arising from the DWBA calculations and the choice of OMPs.
\label{tab:table5}
}
\begin{ruledtabular}
\setlength{\tabcolsep}{2pt}
\begin{tabular}{lllrllr}
 \multirow{2}{*}{Nucleus} &\multicolumn{3}{c}{LE Parameters}&\multicolumn{3}{c}{HE Parameters} \\ \cline{2-4} \cline{5-7}
 &$E_{\mathrm{LE}} $ & $\Gamma_{\mathrm{LE}}$ & EWSR & $E_{\mathrm{HE}} $ & $\Gamma_{\mathrm{HE}}$ & EWSR \\
 &(MeV) &(MeV) & \% & (MeV) & (MeV) & \%\\
 \cline{1-7}
 \\
 \nuc{104}{Pd} & 13.5$\pm$0.2 & 4.7$^{+0.5}_{-0.4}$ & 115$^{+10}_{-8}$& 28.2$^{+1.2}_{-1.3}$ & 9.3$^{+1.8}_{-2.2}$ &  117$^{+5}_{-4}$ \\ \\
 \nuc{106}{Pd} & 13.7$\pm$0.2 & 3.8$^{+0.4}_{-0.3}$ & 106$\pm$8& 27.1$^{+1.8}_{-1.9}$ &  10.1$^{+1.4}_{-2.1}$ & 88$^{+1}_{-8}$ \\ \\
 \nuc{108}{Pd} & 13.2$\pm$0.2 & 4.4$^{+0.6}_{-0.5}$ & 105$^{+10}_{-9}$ & 27.0$^{+1.8}_{-1.7}$ & 9.6$^{+1.7}_{-2.4}$ &98$^{+1}_{-6}$ \\ \\
 \nuc{110}{Pd} & 13.2$\pm$0.2 & 4.3$^{+0.5}_{-0.4}$ & 125$\pm10$ & 26.4$^{+1.8}_{-1.6}$ & 9.6$^{+1.7}_{-2.5}$ &103$^{+3}_{-8}$
\end{tabular}
\end{ruledtabular}
\end{table*}
Figure \ref{fig:ISGQRs} shows the ISGQR strength distributions in the nuclei investigated in this work. Here also, the strength distributions comprise two distinct components. Therefore, two-Lorentzian fits (LE and HE), similar to the case of the ISGDR, were carried out and are shown in the figure. Table \ref{tab:table5} lists the parameters for the LE and HE components, as well as EWSRs for each.

The LE component is well established as the canonical ISGQR response in giant resonance studies \cite{UGarg}. These LE peaks are all well within statistical uncertainties of one another and exhaust between 
95--135\% of the EWSR.

The HE fit is an attempt to quantify the quadrupole overtone previously discussed by Refs. \cite{HUNYADI2008,ABDULLAH2024138852}. The present results provide further evidence for the existence of the overtone in a new region of atomic nuclei. Centroids for this mode are at nearly twice that of the LE mode, which is expected given that it corresponds to a $4\hbar \omega$ excitation. As this is another higher-order compression mode like the ISGMR and the ISGDR, further exploration of this mode could yield another method for refinement of the $K_{\infty}$ and the EoS.

\section*{IV. Theory}\label{IV}
The experimental results for the ISGMR, the primary focus of this work, are compared with calculations based on the self-consistent QRPA+QPVC theory. It has recently been shown that the inclusion of QPVC effects is essential to achieve a unified description of the ISGMR in $^{48}$Ca, the Sn isotopes, and $^{208}$Pb \cite{Z.Z.Li_2023_PRL_ISGMR}; in other words, the ISGMR in all these various nuclei can be consistently described starting from the same Skyrme EDF, associated with a unique value of the incompressibility, only if we move from the QRPA to the QRPA+QPVC level. At this point, it is relevant to check if the Pd isotopes fit within the same picture. In the following, we assume spherical ground-state for these isotopes.

The details of QRPA and QPVC theories can be found  in Refs.~\cite{Z.Z.Li_2024_PRC,L.Guo_2025_EPJA}. In the QRPA+QPVC theory, the coupling between two quasiparticle states and phonons is considered, so that the physical states 
$\vert \nu \rangle$ are expressed as: 
\begin{equation}
	\begin{aligned}
		\mathcal O_\nu^\dagger &= \sum\limits_{a<b} \Big(X_{ab}^{(\nu)} \alpha_a^\dagger \alpha_b^\dagger - Y_{ab}^{(\nu)} \alpha_b \alpha_a  \Big) \\
		&\quad+ \sum\limits_{a<b,n} \Big( X_{abn}^{(\nu)} \alpha_a^\dagger \alpha_b^\dagger Q_n^\dagger - Y_{abn}^{(\nu)} Q_n \alpha_b \alpha_a \Big) , 
	\end{aligned}
\end{equation}
where  $\alpha^\dagger$ ($\alpha$) is the creation (annihilation) operator of quasiparticles $a$ and $b$, $n$ denotes the phonon state obtained by acting with the QRPA creation operator $Q_n^\dagger$ on the ground state, and $X^{(\nu)}$ and $Y^{(\nu)}$ are the forward-going and backward-going amplitudes associated with the QPVC eigenstates $| \nu \rangle$.  These are obtained by solving the energy-dependent QPVC equation in the two-quasiparticle space, while the subtraction method is  used to avoid double-counting. The strength function associated with QPVC is given by: 
	\begin{align}
		S_{\lambda}(E) = -\dfrac{1}{\pi} \textnormal{Im}\sum\limits_{\nu \mu}  \dfrac{\langle 0 | \hat O_{\lambda \mu}| \nu\rangle^2}{E-\Omega_\nu + i\big( \frac{\Gamma_\nu}{2} + \eta \big)}, \label{eq3}
	\end{align}
	where $\hat O_{\lambda\mu}$ is a multipole excitation operator, while $\Omega_\nu - \frac{i}{2}\Gamma_\nu$ are the complex QPVC energies. 

	In the monopole case, 
	\begin{equation}
		\begin{aligned}
			\hat O&= \sum\limits_{i=1}^{A}  r_i^2.
		\end{aligned}
	\end{equation}
	
\begin{figure}[]
\vspace{0.2cm}
	\includegraphics[width=0.45\textwidth]{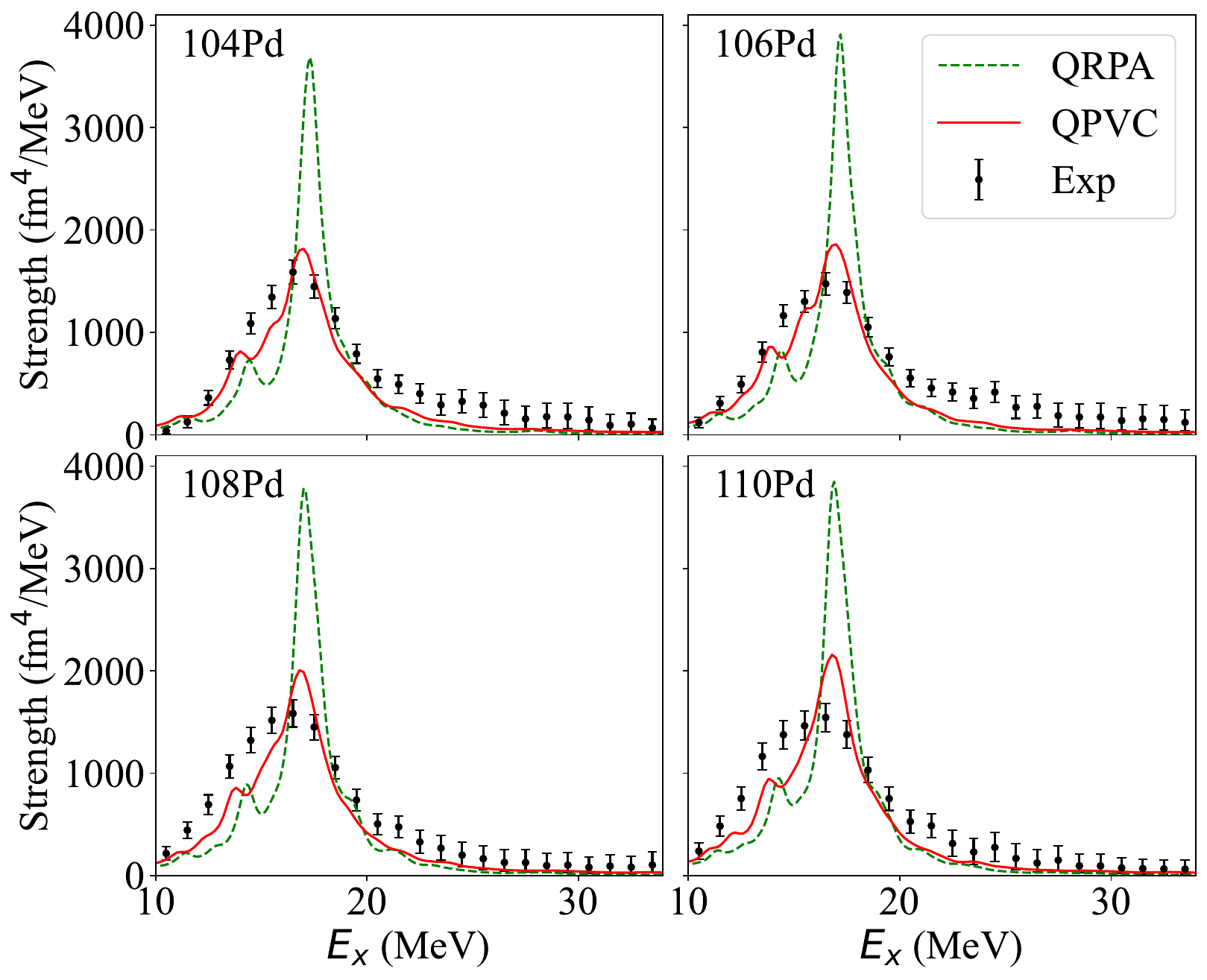} 
	\caption{The ISGMR strength functions in $^{104,106,108,110}$Pd isotopes, calculated either by QRPA using a smoothing with Lorentzian having a width of 1 MeV (dashed green line) or QRPA+QPVC (solid red line). The SV-K226 Skyrme force is used. The experimental data are given by filled black circles.}
	\label{fig-theo}
\end{figure}

To perform the calculations for the Pd isotopes, we adopt the SV-K226 Skyrme interaction in the p-h channel, characterized by a clear value of the incompressibility and a volume pairing force in the p-p channel as in Ref.~\cite{Z.Z.Li_2023_PRL_ISGMR}. The pairing strengths are fitted using the pairing gaps in $^{106}$Pd according to the standard five-point formula. The ISGMR strength functions are shown in Fig.~\ref{fig-theo}. With the inclusion of QPVC effects, the theoretical description is significantly improved. A comparison of the centroid energies from QRPA and QRPA+QPVC calculations with the experimental values, presented in Table \ref{tab:table6}, clearly demonstrates this improved agreement. 

Total widths are also comparable with the experimental ones, though there remains a discrepancy between centroid energies.

\section*{V. Conclusions}\label{V}

Inelastic $\alpha$ scattering measurements off \nuc{104,106,108,110}{Pd} were performed to measure isoscalar giant resonance strengths in these nuclei. The extracted ISGMR strengths in all these isotopes exhibit nominal characteristics compared to previous studies of isotopic chains. Additionally, isobaric comparisons between \nuc{106,110}{Pd} and \nuc{106,110}{Cd} further exemplify the quality of the results obtained here. Leptodermous analysis and the extracted $K_\tau$ value of $-520\pm$140 MeV from this work also agrees well with the values previously obtained in Refs. \cite{TaoLi_116Sn, PATEL2012447}. 

There is good agreement between ISGMR strength distributions extracted here and theoretical calculations utilizing QRPA+QPVC. A pure QRPA approach shows a large centroid discrepancy for all studied nuclei here, but inclusion of QPVC effects dramatically improves the agreement. With QPVC included, there is still a persistent energy difference for all nuclei, although this is less severe when acknowledging the substructure of the strength distribution from QRPA+QPVC calculations. Inclusion of ground-state deformation effects with QPRA+QPVC could further improve the persistent difference in strength distributions.

\begin{table}
\caption{QRPA, QRPA+QVPC, and experimental ISGMR centroid energies for all nuclei investigated in this work. The KV-K226 Skyrme force was used in these calculations.}\label{tab:table6}
\begin{ruledtabular}
\setlength{\tabcolsep}{2pt}
    \begin{tabular}{cccc}
     \multirow{2}{*}{Nucleus}  & QRPA & QRPA+QPVC & Exp.  \\
     & (MeV) & (MeV) & (MeV) \\\hline
     \\ \nuc{104}{Pd} & 17.30 & 16.86 & 16.7$\pm$0.1\\
     \\ \nuc{106}{Pd} & 17.23 & 16.72 & 16.6$\pm$0.1 \\
     \\ \nuc{108}{Pd} & 17.13 & 16.66 & 16.2$\pm$0.1  \\
     \\ \nuc{110}{Pd} & 17.01 & 16.62 & 16.0$\pm$0.2 \\
\end{tabular}
\end{ruledtabular}
\end{table}

Both the ISGDR and the ISGQR also exhibit nominal responses, though the ISGDR has errors larger uncertainty than even multipolarties in this analysis due purely to statistical effects from the MDA algorithm. For the ISGQR in particular, a non-standard higher order Lorentzian was fit to attempt to quantify the quadrupole overtone that has been studied recently \cite{HUNYADI2008,ABDULLAH2024138852}. The centroid of this higher-energy component is roughly double that of the lower-energy centroid, which is indicative of the $4\hbar\omega$ overtone in comparison with the $2\hbar\omega$ ISGQR response.

This work has charted isoscalar giant resonance responses for a new medium-heavy isotopic chain important for the softness conundrum. Modern methods that include QPVC effects into standard QRPA approaches greatly alleviate the differences observed in this work.

\section*{Acknowledgments}

We are grateful to the RCNP cyclotron operators for providing the high-quality $\alpha$ beams required for these measurements. This work was supported in part by the National Science Foundation (Grants No. PHY-2011890 and No. PHY-2310059). Further support was provided by IReNA under OISE-1927130. 

%

\end{document}